# アプリケーションループ文のFPGA自動オフロード手法の評価

山登庸次†

† NTT ネットワークサービスシステム研究所，東京都武蔵野市緑町 3-9-11
E-mail: †yoji.yamato.wa@hco.ntt.co.jp

**あらまし**　近年，ムーアの法則の減速が予測されており，それに伴い特に CPU に比べ電力効率の良い FPGA 等の CPU 以外のハードウェアの活用が増している．しかし，CPU 以外のヘテロなハードウェアの利用には，OpenCL 等の技術スキルの障壁が高い．それを踏まえ，私は，一度書いたコードを，配置されるハードウェアに応じて，自動変換，設定し，高性能で運用できるようにする環境適応ソフトウェアを提案している．GPU へのオフロードについては，一部を自動化してきた．本稿では，FPGA へのオフロードの最初のステップとして，ソースコードの適切なオフロード対象ループ文の自動抽出手法について提案，評価する．提案手法を市中のアプリケーションで有効性を評価する．

# Evaluation of Automatic FPGA Offloading for Loop Statements of Applications

Yoji YAMATO†

† Network Service Systems Laboratories, NTT Corporation, 3-9-11, Midori-cho, Musashino-shi, Tokyo
E-mail: †yoji.yamato.wa@hco.ntt.co.jp

**Abstract**　In recent years, with the prediction of Moore's law slowing down, utilization of hardware other than CPU such as FPGA which is energy effective is increasing. However, when using heterogeneous hardware other than CPUs, barriers of technical skills such as OpenCL are high. Based on that, I have proposed environment adaptive software that enables automatic conversion, configuration, and high-performance operation of once written code, according to the hardware to be placed. Partly of the offloading to the GPU was automated previously. In this paper, I propose and evaluate an automatic extraction method of appropriate offload target loop statements of source code as the first step of offloading to FPGA. I evaluate the effectiveness of the proposed method using an existing application.

## 1. はじめに

近年，CPU の半導体集積度が 1.5 年で 2 倍になるというムーアの法則が減速するのではないかと言われている．そのような状況から，マルチコアの CPU だけでなく，FPGA（Field Programmable Gate Array）や GPU（Graphics Processing Unit）等のハードウェアの活用が増えている．特に FPGA は電力効率が CPU に比べて良く，FPGA の活用により電力削減も期待されている．例えば，Microsoft 社は FPGA を使って Bing の検索効率や DC の電力効率を高めるといった取り組みをしており [1]，Amazon 社は，FPGA, GPU 等をクラウド技術（例えば，[2]- [14]）でインスタンスとして提供している [15].

しかし，CPU 以外のハードウェアをシステムで適切に活用するためには，ハードウェアを意識した設定やプログラム作成が必要であり，OpenCL（Open Computing Language）[16]，CUDA（Compute Unified Device Architecture）[17] といった知識が必要になってくるため，大半のプログラマーにとっては，スキルの壁が高い．

一方，IoT（Internet of Things）技術（例えば，[18]- [22]）は普及してきており，ネットワークにつながるデバイスも，既に数百億と増えており，2030 年には兆台がつながると予測されている．IoT を用いた応用は，医療，流通，製造，農業，エンタメ等に広がっており，サービス連携技術等（例えば，[23]- [30]）を活用して，製品が届くまでの過程を可視化するなどの数多く

の応用がされている．

IoT を用いたシステムで，IoT デバイスを詳細まで制御するためには，組み込みソフトウェア等のスキルが必要になることがある．Raspberry Pi 等の小型端末をゲートウェイ（GW）に，多数のセンサデバイスを集約管理することも頻繁にされるが，小型端末の計算リソースは限定されるため，利用環境に応じて管理の設計が必要となる．

背景を整理すると，CPU 以外の GPU や FPGA 等のハードウェア，多数の IoT デバイスを活用するシステムは今後ますます増えていくと予想されるが，それらを最大限活用するには，壁が高い．そこで，そのような壁を取り払い，CPU 以外のハードウェアや多数の IoT デバイスを十分利用できるようにするため，プログラマーが処理ロジックを記述したソフトウェアを，配置先の環境（FPGA，GPU や IoT GW 等）にあわせて，適応的に変換，設定し，環境に適合した動作をさせるような，プラットフォームが求められている．

Java [31] は 1995 年に登場し，一度記述したコードを，別メーカーの CPU を備える機器でも動作可能にし，環境適応に関するパラダイムシフトをソフト開発現場に起こした．しかし，移行先での性能については，適切であるとは限らなかった．そこで，私は，一度記述したコードを，配置先の環境に存在する GPU や FPGA，IoT GW 等を利用できるように，変換，リソース設定等を自動で行い，アプリケーションを高性能に動作させることを目的とした，環境適応ソフトウェアを提案した．合わせて，環境適応の一要素として，ソフトウェアの GPU へのオフロードの自動化を，ある範囲のアプリケーションに対して実現している [32] [33]．本稿では，FPGA へのオフロードの自動化の第一段として，ソフトウェアのループ文の適切なオフロード領域の自動抽出手法を提案する．提案手法を実装し，市中のアプリケーションで FPGA 自動オフロードの有効性評価をする．

## 2. 既 存 技 術

環境適応ソフトウェアとしては，Java がある．Java は，仮想実行環境である Java Virtual Machine により，一度記述した Java コードを再度のコンパイル不要で，異なるメーカー，異なる OS の CPU マシンで動作させている（Write Once, Run Anywhere）．しかしながら，移行先で，どの程度性能が出るかはわからず，移行先でのデバッグや性能に関するチューニングの稼働が大きい課題があった（Write Once, Debug Everywhere）．

GPU の並列計算パワーを画像処理でないものにも使う GPGPU（General Purpose GPU）（例えば [34]）を行うための環境として CUDA が普及している．CUDA は GPGPU 向けの NVIDIA 社の環境だが，FPGA，メニーコア CPU，GPU 等のヘテロなハードウェアを同じように扱うための仕様として OpenCL が出ており，その開発環境 [35] [36] も出てきている．CUDA，OpenCL は，C 言語の拡張を行いプログラムを行う形だが，プログラムの難度は高い（FPGA 等のカーネルと CPU のホストとの間のメモリデータのコピーや解放の記述を明示的に行う等）

CUDA や OpenCL に比べて，より簡易にヘテロなハードウェアを利用するため，指示行ベースで，並列処理等を行う箇所を指定して，指示行に従ってコンパイラが，GPU 等に向けて実行ファイルを作成する技術がある．仕様としては，OpenACC [37] や OpenMP 等，コンパイラとして PGI コンパイラ [38] や gcc 等がある．OpenACC は，Fortran/C/C++向けの仕様であるが，Java 向けには，IBM の Java JDK [39] が，Java のラムダ記述に従った GPU オフロード処理を行える．

CUDA，OpenCL，OpenACC 等の技術仕様を用いることで，FPGA や GPU へオフロードすることは可能になっている．しかしハードウェア処理自体は行えるようになっても，高速化することには課題がある．例えば，マルチコア，メニーコア CPU 向けに自動並列化機能を持つコンパイラとして，Intel コンパイラ [40] 等がある．これらは，自動並列化時に，コードの中の for 文，while 文等の中で並列処理可能な部分を抽出して，並列化している．しかし，FPGA や GPU を用いる際は，CPU と FPGA，GPU の間のメモリデータ転送のオーバヘッドのため，並列化しても性能がでないことも多い．FPGA や GPU により高速化する際には，OpenCL や CUDA の技術者がチューニングを繰り返したり，PGI コンパイラ等を用いて適切な並列処理範囲を探索し試行することがされている．

このため，OpenCL や CUDA 等の技術スキルが乏しいプログラマーが，FPGA や GPU を活用してソフトウェアを高速化することは難しいし，自動並列化技術等を使う場合も並列処理箇所の試行錯誤等の稼働が必要だった．

## 3. ループ文の FPGA 自動オフロード手法の提案

### 3.1 環境適応処理のフロー

ソフトウェアの環境適応を実現するため，図 1 の処理フローを提案している．環境適応ソフトウェアは，環境適応機能を中心に，検証環境，商用環境，テストケース DB，コードパターン DB，設備リソース DB の機能群が連携することで動作する．

Step1 コード分析：

Step2 オフロード可能部抽出：

Step3 適切なオフロード部探索：

Step4 リソース量調整：

Step5 配置場所調整：

Step6 実行ファイル配置と動作検証：

Step7 運用中再構成：

ここで，Step 1-7 で，環境適応するために必要となる，コードの変換，リソース量の調整，配置場所の決定，Jenkins 等 [41] [42] 使った検証，運用中の再構成を行うことができるが，実施したい処理だけ切り出すこともできる．例えば，本稿で対象とする FPGA 向けのコード変換だけ実施する場合は，Step 1-3 だけ処理すればよく，利用する機能も環境適応機能や検証環境等だけ利用する形でよい．

### 3.2 FPGA 自動オフロードに向けた考慮事項

Step 1 のコード分析処理では，Clang [43] 等の市中の構文解析ツールを用いて，ソースコードのパース分析を行う．コード分析は，オフロードする移行先デバイスを想定した分析が必要

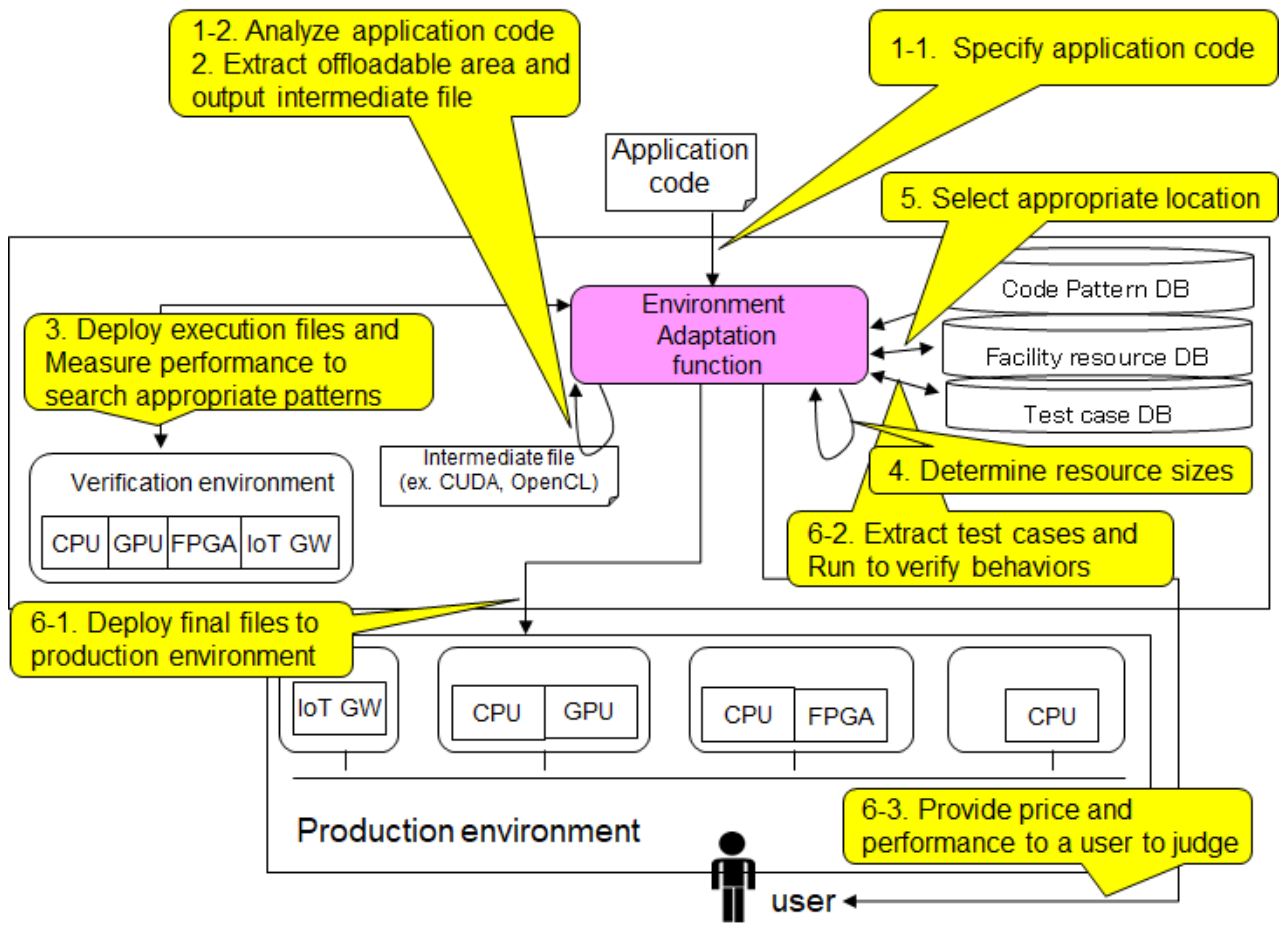

図 1　環境適応ソフトウェアのフロー

になると考えるため，一般化することは難しいが，変数参照関係やループ文等のコード構造のプリミティブな把握と，例えばフーリエ変換処理を行う機能である事やフーリエ変換処理を行うライブラリを呼び出している等の機能ブロック利用を把握する．利用している機能ブロックの分析は，機械が自動判断することは極めて難しいが，Deckard 等の類似コード検出ツール等を活用して類似度等で分析することを考えている．なお，Clang は C 言語，C++向けツールであるが，解析する言語に合わせて市中ツールを選ぶ必要がある．

Step 2 及び 3 は CPU 処理をオフロードする，FPGA，GPU，IoT GW 等の，オフロード先に合わせた検討が必要であるため，個々のオフロード先毎に処理機能をプラグインする形を想定する．ただし，一般的に，性能は，最大性能になる設定を自動発見する事は難しいため，オフロードするパターンを準備し，性能測定を検証環境で試行し，より高速化できるパターンを見つける方針をとる．GPU については，[32] で，進化的計算手法である遺伝的アルゴリズム（GA）[44] を用いてループ回数の多いループ文の自動オフロードの実現性を確認している．そこで，本稿では，FPGA 向けオフロード手法について検討する．

オフロードしたいアプリケーションは，多様であるが，映像処理のための画像分析（[45] [46] 等），センサデータを分析するための機械学習処理等，計算量，時間が多いアプリケーションでは，ループ文による繰り返し処理が長時間を占めている．そこで，本稿では，ループ文を FPGA に自動でオフロードする事での高速化を狙う．

しかし，高速化にはオフロードするコードの適切なパイプライン処理，並列処理等が必要である．特に，FPGA を使う場合は，CPU と FPGA 間のデータ転送が生じるため，データのサイズやループの回数が大きくないと性能が出ない．FPGA のクロック速度も CPU より遅いため，FPGA に適した計算でないと性能が出ない．また，メモリプロセス持ち方やメモリのデータ転送タイミング等によっては，FPGA で高速化できる個々のループ文の組み合わせが，最速とならない場合も考えられる（例：5 個の for 文で，1 番，3 番，5 番の 3 つが CPU に比べて高速に FPGA で処理出来る場合に，1 番，3 番，5 番の 3 つの組み合わせが最速になるとは限らない）．

GPU の場合は，OpenACC を用いて#pragma acc kernels のディレクティブ記述で，指定したループ文の GPU 実行をしたり，CUDA でより細かい指定が出来た．FPGA では，OpenCL を用いた指定や，より抽象的な高位合成ツールを用いた FPGA 処理の指定が可能である．OpenCL では，以下の 10 ステップの記載が必要である．デバイスの準備，カーネルの準備，デバイスメモリの割り当て，ホストからデバイス方向のデータ転送，カーネル関数の引数設定，カーネル関数の実行，デバイスから方向のデータ転送，デバイスメモリの解放，カーネルの解放，その他オブジェクト（デバイス）の解放．高位合成ツールは提供ベンダにより仕様は大きく異なるが，例えば，Xilinx 社の vivado HLS では，OpenACC と類似の#pragma HLS UNROLL 等のディレクティブで FPGA 処理を指定できる．Intel FPGA SDK for OpenCL では，OpenCL 標準に加え，#pragma ディレクティブで指定もできるよう拡張している．

以前の研究である [32] は，CPU 向けの汎用的コードから，GPU オフロードに向けて，適切な並列処理領域を自動抽出するため，並列処理可能なループ文群に対して GA を用いて，より適切な並列処理領域を探索する事を，GPU を備える検証環境で性能検証を反復しながら行っている．しかし，FPGA では，プログラムサイズやマシンスペックにもよるが，コンパイルして FPGA 実機で動作させるのに数時間程度かかるのが一般的である．そのため，[32] のように GA 等で，多数のパターンの性能測定を繰り返し行うのは現実的でない．そこで，実際に性能測定するパターンを絞ってから検証環境に配置し，コンパイルして FPGA 実機で性能測定する回数を減らすことを想定する．

### 3.3　FPGA 自動オフロード手法

前サブ節の考慮事項を踏まえて，ループ文の FPGA 自動オフロード手法を提案する．

本手法は，まず，オフロードしたいソースコードの分析を行う．前サブ節記載の通り，ソースコードの言語に合わせて，ループ文や変数の情報を把握する．

次に，把握したループ文に対して，FPGA オフロードを試行するかどうか候補を絞っていく処理を行う．ループ文がオフロード効果があるかどうかは，算術強度が一つの指標となりうる．算術強度は，ループ回数やデータ量が多いと増加し，アクセス数が多いと減少する指標であり，算術強度が高い処理はプロセッサにとって重い処理となり時間がかかる．そこで，算術強度分析ツールで，ループ文の算術強度を分析し，密度が高いループ文をオフロード候補に絞る．

高算術強度のループ文であっても，それを FPGA で処理する際に，FPGA リソースを過度に消費してしまうのは問題である．そこで次に，高算術強度ループ文を FPGA 処理する際のリソース量の算出に移る．FPGA にコンパイルする際の処理としては，OpenCL 等の高位言語からハードウェア記述の HDL 等のレベルに変換され，それに基づき実際の配線処理等がされる．この時，配線処理等は多大な時間がかかるが，HDL 等の途中状態の段階までは時間は分単位でしかかからない．HDL 等

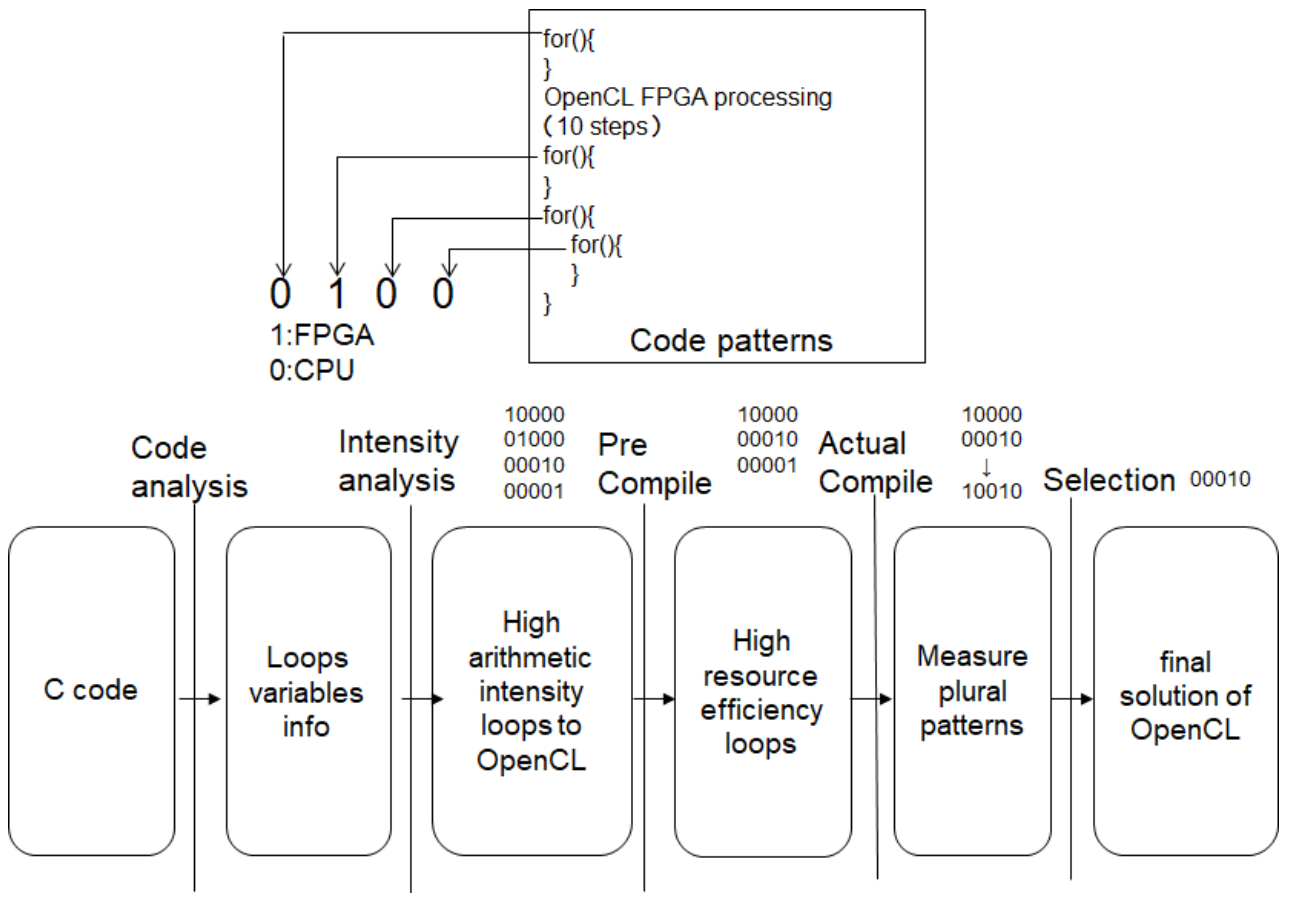

図 2　ループ文の FPGA 自動オフロード手法

のレベルで，FPGA で利用する Flip Flop や Look Up Table 等のリソースは分かるため，利用するリソース量はコンパイルが終わらずとも短時間でわかる．対象のループ文を OpenCL 等の高位言語化し，リソース量を算出することで，ループをオフロードした際の算術強度とリソース量が決まるため，算術強度/リソース量をリソース効率とする．本手法では，高リソース効率のループ文をオフロード候補として更に絞り込む．

ここで，ループ文を OpenCL 等の高位言語化する際には，2 つ処理が必要である．一つは，CPU 処理のプログラムを，カーネル（FPGA）とホスト（CPU）に，OpenCL 等の高位言語の文法に従って分割することである．もう一つは，分割する際に，高速化するための技法を盛り込むことである．一般に，FPGA を用いて高速化するためには，ローカルメモリキャッシュ，ストリーム処理，複数インスタンス化，ループ文の展開処理，ネストループ文の統合，メモリインターリーブ等がある．これらは，ループ文によっては，絶対効果があるわけではないが，高速化するための手法として，よく利用されている．

次に，高リソース効率のループ文が幾つか選択されたため，それらを用いて性能を実測するオフロードパターンを作成する．FPGA での高速化は，1 個の処理だけ FPGA リソース量を集中的にかけて高速化する形もあれば，複数の処理に FPGA リソースを分散して高速化する形もある．まず，選択された単ループ文に対してパターンを作って FPGA 実機で動作するようコンパイルして性能測定する．次に，高速化できる単ループ文に対してはその組み合わせのパターンも 2 回目に作り，同様に性能測定する．

最後に，検証環境で性能測定された複数パターンの中で，高速のパターンを解として選択する．

このように，ループ文の FPGA 自動オフロードは，算術強度が高くリソース効率が高いループ文に絞って，オフロードパターンを作り，検証環境で実測を通じて高速なパターンの探索を行う（図 2）．

## 4. 実　　装

3 節提案技術の有効性を確認するための実装を説明する．FPGA 自動オフロードの有効性確認のため，対象アプリケーションは C/C++言語のアプリケーションとし，FPGA は Intel PAC with Intel Arria10 GX FPGA を用いる．なお，コンパイルするマシンは，DELL EMC PowerEdge R740（CPU：Intel Xeon Bronze 3104 / 1.70GHz，RAM：32GB RDIMM DDR4-2666 × 2）である．

FPGA 処理は，Intel Acceleration Stack Version 1.2（Intel FPGA SDK for OpenCL 17.1.1，Quartus Prime Version 17.1.1）を用いる．Intel FPGA SDK for OpenCL は，標準 OpenCL に加え Intel 向けの#pragma 等を解釈する高位合成ツール（HLS）であり，FPGA で処理するカーネルと CPU で処理するホストプログラムを記述した OpenCL コードを，解釈しリソース量等の情報を出力し，FPGA の配線作業等を行い，FPGA で動作できるようにする．FPGA 実機で動作できるようにするには，100 行程度の小プログラムでも 3 時間程の長時間がかかるが，リソース量オーバーの際は早めにエラーとなる．また，FPGA で処理できない OpenCL コードの際は，数時間後にエラーを出力する．

実装の動作概要を説明する．言語は python 2.7 で行い，以下の処理を行う．

実装は，C/C++アプリケーションの利用依頼があると，まず，C/C++アプリケーションのコードを解析して，for 文を発見するとともに，for 文内で使われる変数データ等の，プログラム構造を把握する．構文解析には，LLVM/Clang の構文解析ライブラリ (libClang の python binding) を使う．

実装は，次に，各ループ文の FPGA オフロード効果があるかの見込みを得るため，算術強度分析ツールを実行し，ループ回数，データサイズ，アクセス数等で定まる算術強度の指標を取得する．算術強度上位 a 個のループ文のみ対象とするようにする．算術強度分析には，PGI コンパイラ 19.4 を用いる．なお，PGI コンパイラは GPU 向けのコンパイラであるが，算術強度分析には利用できるので，算術強度分析部分だけ利用する (gcov [47]，gprof 等もループ回数算出には利用できる)．

実装は，次に，高算術強度の個々のループ文に対して，FPGA オフロードする OpenCL コードを生成する．OpenCL コードは当該ループ文を FPGA カーネルとして，残りを CPU ホストプログラムとして分割したものである．FPGA カーネルコードとする際に，高速化の技法としてループ文の展開処理を b だけ行う．ループ文展開処理は，リソース量は増えるが，高速化に効果があり，展開する数は一定数 b に制限してリソース量が膨大にならない範囲で行う．

実装は，次に，a 個の OpenCL コードに対して，Intel FPGA SDK for OpenCL を用いて，プレコンパイルをして，利用する Flip Flop，Look Up Table 等のリソース量を算出する．使用リソース量は全体リソース量の割合で表示される．ここで算術強度とリソース量から，各ループ文のリソース効率を計算する．例えば算術強度が 10，リソース量が 0.5 のループ文は 10/0.5=20，算術強度が 3，リソース量が 0.3 のループ文は 3/0.3=10 がリソース効率となり，前者が高い．各ループ文で，リソース効率が高い c 個を選定する．

実装は，次に，c 個のループ文を候補に，実測するパターンを作る．例えば，1 番目と 3 番目と 5 番目のループが高リソース効率であったとしたら，まず，1 番をオフロード，3 番をオフロード，5 番をオフロードの OpenCL パターンを作成しコンパイルする．最初に実測するパターンは d 個以内で作成し，検証環境の FPGA を備えたサーバで性能測定を行う．性能測定には高速化したいアプリケーションで指定されたサンプル処理を行う．例えば，フーリエ変換のアプリケーションであれば，サンプルデータでの変換処理をベンチマークに性能測定をする．この中で，1 番と 3 番のオフロードで効果があった場合は，2 回目は 1 番と 3 番の両方をオフロードする OpenCL パターンを作成し，コンパイルする．なお，ループの組み合わせを作る際は，利用リソース量も組み合わせになるため上限値に納まらない場合は，その組合せパターンは作らない．

実装は，最後に，検証環境で測定された複数のパターンから高速なパターンを解として選択する．

## 5. 評　　価

### 5.1 評 価 条 件

#### 5.1.1 評 価 対 象

評価対象は，信号処理の時間領域有限インパルス応答フィルタと画像処理の MRI-Q とする．

時間領域有限インパルス応答フィルタは，システムにインパルス関数を入力したときの出力に対して有限時間で打ち切る処理を行う，フィルタの一種である．実装は種々あるが，[48] の C コードを用いて，性能測定に用いるサンプルテストも備え付けのテストを用いる．IoT 等で，デバイスからの信号データをネットワーク転送するアプリケーションを考えた際に，ネットワークコストを下げるため，フィルタ等の信号処理をして送ることは想定される．そのため，信号処理の FPGA での自動高速化は応用範囲が広いと考える．

MRI-Q [49] は，MRI (Magnetic Resonance Imaging) 3D 画像を処理するプログラムである．IoT 領域では，カメラを用いた自動検知等のため，画像処理は頻繁に利用されるため，高速化のニーズは強い．

#### 5.1.2 評 価 手 法

GPU でのオフロード検討 [33] と異なり，FPGA では数多くのパターンで性能測定は行わない．検証では，検証対象アプリケーションループ文の，算術強度，リソース効率，コンパイル中の HDL 相当情報等の途中情報と共に，検証環境での複数のオフロードパターンのサンプルアプリでの性能測定結果を記録している．性能測定を通じて，最高性能パターンが，オフロード探索の解であり，その解の性能を，全て CPU 処理に比べて，改善度を評価する．

実験の条件は以下で行う．

オフロード対象：ループ文数（時間領域有限インパルス応答フィルタは 36．MRI-Q は 16．）

算術強度絞り込み：算術強度分析の上位 5 つのループ文に絞り込み

ループ文展開数：1（展開処理や複数インスタンス化はリソース量を使うほど高速化できるケースが多いため，検証では OpenCL での FPGA オフロードした効果だけ確認するため）

リソース効率絞り込み：リソース効率分析の上位 3 つのループ文に絞り込み（算術強度/リソース量が高い上位 3 つのループ文を選定）

実測オフロードパターン数：4（1 回目は上位 3 つのループ文オフロードパターンを測定し，2 回目は 1 回目で高性能だった 2 つのループ文オフロードの組み合わせパターンで測定．）

#### 5.1.3 評 価 環 境

検証 FPGA として Intel PAC with Intel Arria10 GX FPGA を用いる．FPGA 処理は，Intel Acceleration Stack Version 1.2 を用いる．評価環境とスペックを図 3 に示す．ここで，Client ノート PC から，ユーザが利用する C/C++言語アプリケーションコードを指定し，Verification Machine を用いてチューニング後，実際のユーザが使う Running environment にデプロイする形である．

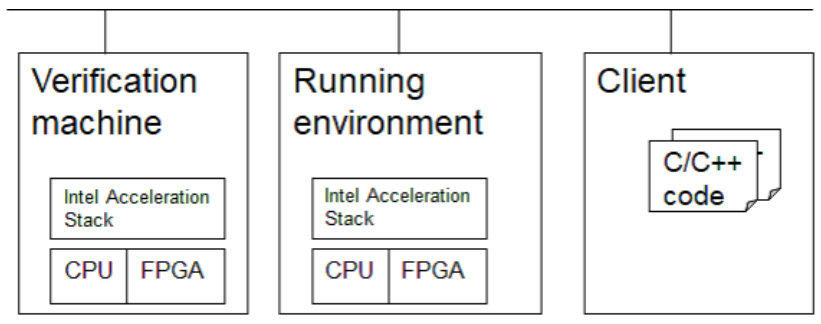

| Name | Hardware | CPU | RAM | FPGA | OS | Intel Acceleration Stack |
|---|---|---|---|---|---|---|
| Verification machine | Dell PowerEdge R740 | Intel Xeon Bronze 3104 | 32GB*2 | Intel PAC with Intel Arria10 GX FPGA | CentOS 7.4 | 1.2 |
| Running environment | Dell PowerEdge R740 | Intel Xeon Bronze 3104 | 32GB*2 | Intel PAC with Intel Arria10 GX FPGA | CentOS 7.4 | 1.2 |
| Client | HP ProBook 470 G3 | Intel Core i5-6200U @2.3GHz | 8GB | | Windows 7 Professional | |

図 3　性能測定環境

| | Performance improvement of this paper implementation |
|---|---|
| Time domain finite impulse response filter | 4.0 |
| MRI-Q | 7.1 |

図 4　提案 FPGA 自動オフロード手法での性能改善度

### 5.2 性 能 結 果

FPGA で手動高速化が良く行われるアプリケーションとして，時間領域有限インパルス応答フィルタ，MRI-Q での高速化を確認した．

今回の FPGA 自動化の提案手法の実装により，どの程度性能が改善されたかの測定結果を図 4 に示す．図 4 は，最終解の性能が，全 CPU 処理に比べて何倍になっているかを示した表である．図 4 より，時間領域有限インパルス応答フィルタについては，今回提案手法により 4.0 倍の性能が実現できていることがわかる．MRI-Q については，今回提案手法により 7.1 倍の性能が実現できていることがわかる．自動化時間に関しては，大きく時間がかかるのは FPGA のコンパイル時間であり，一つのオフロードパターンのコンパイルに 3 時間ほどかかっているため，4 パターンの検証を自動で行う事で約半日かかっている．

## 6. ま と め

本稿では，私が提案している，ソフトウェアを配置先環境に合わせて自動適応させFPGA 等を適切に利用して，アプリケーションを高性能に運用するための環境適応ソフトウェアの要素として，ソフトウェアループ文を電力効率の良い FPGA へ自動オフロードする手法を提案し，評価した．

提案した FPGA オフロードの自動化手法は，コードを分析して for 等のループ文を発見するまでは，GPU での手法 [32] と同様であるが，FPGA 実機へのコンパイルは長時間かかる現状の対応のため，オフロード候補の for 文を絞ってから，実測試行を行う．発見されたループ文に対して，算術強度分析ツールを用いて算術強度が高いループ文を抽出し，更に，高算術強度のループ文に対して，展開処理等の FPGA オフロード化を行う OpenCL をプレコンパイルして，リソース効率が高いループ文を見つけ，更に対象に絞り込む．絞り込まれたループ文に対して，個々のループ文をオフロードした OpenCL やそれらループ文を組み合わせた OpenCL を生成し，FPGA 実機へコンパイルして，性能測定を行い，高速の OpenCL を解として選択する．

市中アプリケーションに対して提案手法での FPGA 自動オフロードを行い，方式の有効性を確認した．今後は，多数のアプリケーションで評価を行うとともに，ループ文だけでなく，FFT 等大きな機能ブロック単位でのオフロードも検討する．

### 文 献

[1] A. Putnam, et al., "A reconfigurable fabric for accelerating large-scale datacenter services," ISCA’14, pp.13-24, 2014.
[2] O. Sefraoui, et al., "OpenStack: toward an open-source solution for cloud computing," International Journal of Computer Applications, Vol.55, 2012.
[3] Y. Yamato, et al., "Fast and Reliable Restoration Method of Virtual Resources on OpenStack," IEEE Transactions on Cloud Computing, Sep. 2015.
[4] Y. Yamato, et al., "Evaluation of Agile Software Development Method for Carrier Cloud Service Platform Development," IEICE Transactions on Information & Systems, Vol.E97-D, No.11, pp.2959-2962, Nov. 2014.
[5] Y. Yamato, "Automatic verification technology of software patches for user virtual environments on IaaS cloud," Journal of Cloud Computing, Springer, 2015, 4:4, Feb. 2015.
[6] Y. Yamato, "Optimum Application Deployment Technology for Heterogeneous IaaS Cloud," Journal of Information Processing, Vol.25, No.1, pp.56-58, Jan. 2017.
[7] Y. Yamato, "Performance-Aware Server Architecture Recommendation and Automatic Performance Verification Technology on IaaS Cloud," Service Oriented Computing and Applications, Springer, Nov. 2016.
[8] Y. Yamato, "Server Selection, Configuration and Reconfiguration Technology for IaaS Cloud with Multiple Server Types," Journal of Network and Systems Management, Springer, Aug. 2017.
[9] Y. Yamato, et al., "Development of Low User Impact and Low Cost Server Migration Technology for Shared Hosting Services," IEICE Transactions on Communications, Vol.J95-B, No.4, pp.547-555, Apr. 2012.
[10] Y. Yamato, et al., "Development of Resource Management Server for Production IaaS Services Based on OpenStack," Journal of Information Processing, Vol.23, No.1, pp.58-66, Jan. 2015.
[11] Y. Yamato, et al., "Software Maintenance Evaluation of Agile Software Development Method Based on OpenStack," IEICE Transactions on Information & Systems, Vol.E98-D, No.7, pp.1377-1380, July 2015.
[12] Y. Yamato, "Cloud Storage Application Area of HDD-SSD Hybrid Storage, Distributed Storage and HDD Storage," IEEJ Transactions on Electrical and Electronic Engineering, Vol.11, Issue.5, pp.674-675, Sep. 2016.
[13] Y. Yamato, "Use case study of HDD-SSD hybrid storage, distributed storage and HDD storage on OpenStack," 19th International Database Engineering & Applications Symposium (IDEAS15), pp.228-229, July 2015.
[14] Y. Yamato, "OpenStack Hypervisor, Container and Baremetal Servers Performance Comparison," IEICE Communication Express, Vol.4, No.7, pp.228-232, July 2015.
[15] AWS EC2 web site, https://aws.amazon.com/ec2/instance-types/
[16] J. E. Stone, et al., "OpenCL: A parallel programming standard for heterogeneous computing systems," Computing in science & engineering, Vol.12, No.3, pp.66-73, 2010.
[17] J. Sanders and E. Kandrot, "CUDA by example : an introduction to general-purpose GPU programming," Addison-Wesley, 2011
[18] M. Hermann, et al., "Design Principles for Industrie 4.0 Scenarios," Rechnische Universitat Dortmund. 2015.
[19] Y. Yamato, "Experiments of posture estimation on vehicles using wearable acceleration sensors," The 3rd IEEE International Conference on Big Data Security on Cloud (BigDataSecurity 2017), pp.14-17, May 2017.
[20] Y. Yamato, et al., "Predictive Maintenance Platform with Sound Stream Analysis in Edges," Journal of Information Processing, Vol.25, pp.317-320, Apr. 2017.
[21] Tron project web site, http://www.tron.org/
[22] P. C. Evans and M. Annunziata, "Industrial Internet: Pushing the Boundaries of Minds and Machines," Technical report of General Electric (GE), Nov. 2012.
[23] Y. Yamato, "Ubiquitous Service Composition Technology for Ubiquitous Network Environments," IPSJ Journal, Vol.48, No.2, pp.562-577, Feb. 2007.
[24] Y. Yamato, et al., "Study of Service Processing Agent for Context-Aware Service Coordination," IEEE International Conference on Service Computing (SCC 2008), pp.275-282, July 2008.
[25] Y. Yamato, et al., "Study and Evaluation of Context-Aware Service Composition and Change-Over Using BPEL Engine and Semantic Web Techniques," IEEE Consumer Communications and Networking Conference (CCNC 2008), pp.863-867, Jan. 2008.
[26] H. Sunaga, et al., "Service Delivery Platform Architecture for the Next-Generation Network," ICIN2008, 2008.
[27] Y. Yokohata, et al., "Service Composition Architecture for Programmability and Flexibility in Ubiquitous Communication Networks," IEEE International Symposium on Applications and the Internet Workshops (SAINTW’06), 2006.
[28] Y. Nakano, et al., "Effective Web-Service Creation Mechanism for Ubiquitous Service Oriented Architecture," The 8th IEEE International Conference on E-Commerce Technology and the 3rd IEEE International Conference on Enterprise Computing, E-Commerce, and E-Services (CEC/EEE 2006), pp.85, June 2006.
[29] Y. Yamato, et al., "Study of Service Composition Engine Implemented on Cellular Phone," Information technology letters, Vol.4, pp.269-271, Aug. 2005.
[30] Y. Yamato, et al., "Development of Service Control Server for Web-Telecom Coordination Service," IEEE ICWS 2008, pp.600-607, Sep. 2008.

[31] J. Gosling, et al., "The Java language specification, third edition," Addison-Wesley, 2005. ISBN 0-321-24678-0.
[32] Y. Yamato, et al., "Automatic GPU Offloading Technology for Open IoT Environment," IEEE Internet of Things Journal, Sep. 2018.
[33] Y. Yamato, "Study of parallel processing area extraction and data transfer number reduction for automatic GPU offloading of IoT applications," Journal of Intelligent Information Systems, Springer, DOI:10.1007/s10844-019-00575-8, 2019.
[34] K. Shirahata, et al., "Hybrid Map Task Scheduling for GPU-Based Heterogeneous Clusters,"IEEE CloudCom, 2010.
[35] Altera SDK web site, https://www.altera.com/products/design-software/embedded-software-developers/opencl/documentation.html
[36] Xilinx SDK web site, https://japan.xilinx.com/html_docs/xilinx2017_4/sdaccel_doc/lyx1504034296578.html
[37] S. Wienke, et al., "OpenACC-first experiences with real-world applications," Euro-Par Parallel Processing, 2012.
[38] M. Wolfe, "Implementing the PGI accelerator model," ACM the 3rd Workshop on General-Purpose Computation on Graphics Processing Units, pp.43-50, Mar. 2010.
[39] K. Ishizaki, "Transparent GPU exploitation for Java," CANDAR 2016, Nov. 2016.
[40] E. Su, et al., "Compiler support of the workqueuing execution model for Intel SMP architectures," In Fourth European Workshop on OpenMP, Sep. 2002.
[41] Jenkins web site, https://jenkins.io/
[42] Selenium web site, https://www.seleniumhq.org/
[43] Clang website, http://llvm.org/
[44] J. H. Holland, "Genetic algorithms," Scientific american, Vol.267, No.1, pp.66-73, 1992.
[45] OpenCV web site, http://opencv.org/
[46] imageJ web site, https://imagej.nih.gov/ij/docs/concepts.html
[47] gcov website, http://gcc.gnu.org/onlinedocs/gcc/Gcov.html
[48] Time Domain Finite Impulse Response Filter web site, http://www.omgwiki.org/hpec/files/hpec-challenge/tdfir.html
[49] MRI-Q website, http://impact.crhc.illinois.edu/parboil/parboil.aspx